\documentstyle[multicol,aps,epsf,amssymb]{revtex}

\def\Pr   {{\rm Pr}}

\def\hmu   {h_\mu}

\def\EE   {{\bf E} }
\def\SI   {{\bf S} }

\def\TI   {{\bf T} }
\def\l2   {{\rm log}_2}

\newtheorem{theorem}{Theorem}

\newtheorem{proposition}{Proposition}

\begin{document}

%\draft command makes pacs numbers print
%\draft

\title{Synchronizing to the Environment:\\
Information Theoretic Constraints on Agent Learning}

\author{James P. Crutchfield}
\address{Santa Fe Institute, 1399 Hyde Park Road, Santa Fe, NM 87501\\
Electronic Address: chaos@santafe.edu}

\author{David P. Feldman}
\address{College of the Atlantic, 105 Eden St., Bar Harbor, ME 
04609\\and Santa Fe Institute, 1399 Hyde Park Road, Santa Fe, NM 87501\\
Electronic Address: dpf@santafe.edu}

\date{\today}
\maketitle

\bibliographystyle{unsrt}

% ************************* ABSTRACT *************************
\begin{abstract}

We show that the way in which the Shannon entropy of sequences produced
by an information source converges to the source's entropy rate can be
used to monitor how an intelligent agent builds and effectively uses
a predictive model of its environment. We introduce natural measures
of the environment's apparent memory and the amounts of information
that must be (i) extracted from observations for an agent to synchronize
to the environment and (ii) stored by an agent for optimal prediction.
If structural properties are ignored, the missed regularities are
converted to apparent randomness. Conversely, using representations
that assume too much memory results in false predictability.

% Insert PACS numbers on next line
%  old: PACS: 02.50.Wp, 05.45, 05.65+b, 89.70.+c
%02.50.Ga  %Markov processes 
%05.20.-y  %Classical statistical mechanics
05.45.-a  %Nonlinear dynamics and nonlinear dynamical systems
05.45.Tp  %Time series analysis
%65.40.Gr  %Thermodynamics of solids: Entropy and other thermodynamical
	  %quantities  
%89.70.+c  %Information science 
89.75.Kd  %Complex Systems: Patterns 
~~~~~~~~~~Santa Fe Institute Working Paper 01-03-020

\end{abstract}

%\end{frontmatter}

% ****************************************************************

\begin{multicols}{2}

%\tableofcontents
%\newpage 

%  ************************* INTRODUCTION *************************

\section{Untangling Environmental Structure from Randomness}

We examine ways to untangle the different mechanisms responsible for
apparent randomness observed by an intelligent agent. For our purposes
here ``intelligent agent'' simply refers to an observer that (i)
actively builds internal models of its environment using available
sensory stimuli and (ii) takes action based on these models. In
addressing the issue of distinguishing different sources
of environmental noise and structure, we analyze those aspects of
apparent randomness over which an intelligent agent may exercise
control.  These choices include the amount of data to collect, as well
as the choice of statistic or modeling representation used to quantify
the degree of randomness.

% **********************************************************************
\begin{figure}[tbp]
%\vspace{-0.2in}
%\hspace{0.2in}
\epsfxsize=3.0in
\begin{center}
\leavevmode
\epsffile{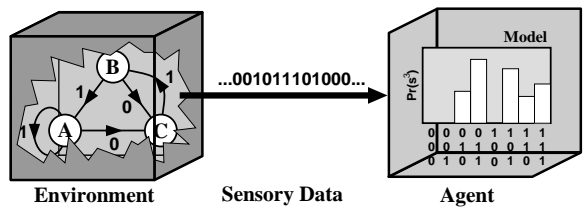}
\end{center}
\caption{The measurement channel: The internal states
$\{{\bf A}, {\bf B}, {\bf C} \}$ of the system are reflected, only
indirectly, in the observed measurement of $1$s and $0$s. An observer
works with this impoverished data to build a model of the underlying
system. After Ref.~\protect\cite{Crut91b}.}
\label{MeasurementChannel}
\end{figure}
% **********************************************************************

One of the central questions addressed in the following is, How does
an agent, apprised of the environment's possible states and behaviors,
come to know the state of its environment? We will show that this
is related to another question, How does an agent come to accurately
estimate how random an environment is? In particular, we investigate
how finite-data approximations converge to an asymptotic measure of
randomness by introducing several quantities that capture the nature of
this convergence.  We demonstrate that regularities that are unseen
are ``converted'' to apparent randomness.  A more thorough discussion
of these results, including proofs of the following propositions
and theorems, is found in \cite{Crut01a}.

% **********************************************************************
\section{Measurement Channel}
\label{Measurement_Channel}

We adapt Shannon's conception of a communication channel as follows:
We assume that there is an {\em environment} (source or process) that
produces a {\em sensory data stream} (message)---a string of symbols
drawn from a finite alphabet ($\cal A$). The task for the
{\em agent} (receiver or observer) is to estimate the probability
distribution of sequences and, thereby, estimate how random the
environment is. Further, we assume that the agent does not
know the environment's structure; the range of its states and their
transition structure---the environment's internal dynamics---are hidden
from the agent.  (We will, however, relax this assumption in
Sec.~\ref{T_Section} below.)  Since the agent does not have direct
access to the environment's internal, hidden states, we picture
instead that the agent simply collects blocks of measurements from the
data stream and stores the block probabilities in a histogram (the
{\em internal model}). In this scheme, the agent can estimate, to
arbitrary accuracy, the probability of measurement sequences by
observing for arbitrary lengths of time.

This {\em measurement channel} scenario is illustrated in
Fig.~\ref{MeasurementChannel}.  In this case, the environment is a
three-state deterministic finite automaton. However, the agent does 
not see the internal states $\{ {\bf A}, {\bf B}, {\bf C} \}$.
Instead, it has access only to the measurement symbols $\{ 0, 1 \}$
generated on state-to-state transitions by the hidden automaton.
In this sense, the measurement channel acts like a communication
channel; the channel maps from an internal-state sequence $\ldots {\bf
BCBAACBC} \ldots$ to a measurement sequence $\ldots 0111010 \ldots$.
The environment depicted in Fig.~\ref{MeasurementChannel} belongs to the
class of stochastic process known as {\em hidden Markov models}.  The
transitions from internal state to internal state are Markovian, in
that the probability of a given transition depends only upon which
state the process is currently in. However, these internal states are
not seen by the agent---hence the name {\em hidden} Markov model
\cite{Blac57a,Elli95}.

%\vspace{-0.5cm}

% **********************************************************************
\section{Entropies: Measuring Randomness}

Let ${\rm Pr}(s^L)$ denote the probability distribution over blocks
$s^L = s_0 , s_1 , \ldots , s_{L-1}$ of $L$ consecutive environment
observations, $s_i \in {\cal A}$. Then the {\em total Shannon entropy}
of these $L$ consecutive measurements is defined to be:
\begin{equation}
H(L) \, \equiv \, - \sum_{ s^L \in {\cal A}^L } \Pr (s^L) \l2 \Pr
(s^L) \;, 
\end{equation}
where $L > 0$.  The sum is understood to run over all possible blocks
of $L$ consecutive symbols. The units of $H(L)$ are {\em bits}. The
entropy $H(L)$ measures the uncertainty associated with sequences of
length $L$.  (For a more detailed discussion of the Shannon entropy and
related information theoretic quantities, see, e.g.,
Ref.~\cite{Cove91}.)  Below, we will focus on the behavior
of the Shannon entropy curve $H(L)$.  We shall see that examining how
$H(L)$ grows with $L$ leads to several quantities that capture aspects
of the environment's randomness and structure.

% **********************************************************************
\begin{figure}[tbp]
%\vspace{-0.2in}
%\hspace{0.2in}
\epsfxsize=2.4in
\begin{center}
\leavevmode
\epsffile{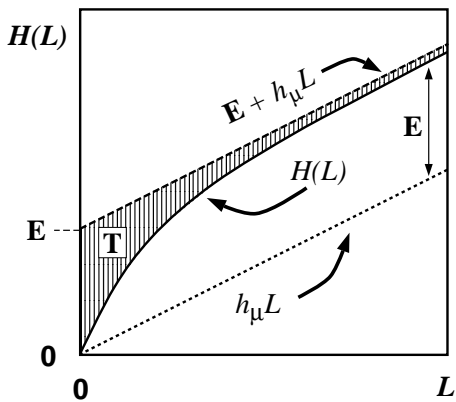}
\end{center}
\caption{Total Shannon entropy growth for a finitary information
source: a schematic plot of $H(L)$ versus $L$. $H(L)$ increases
monotonically and asymptotes to the line $\EE + h_\mu L$, where $\EE$
is the excess entropy and $h_\mu$ is the source entropy rate. 
%This
%dashed line is the $\EE$-memoryful Markovian source approximation to a
%source with entropy growth $H(L)$. The entropy growth of the
%memoryless-source approximation of the source is indicated by the
%short-dashed line $\hmu L$ through the origin with slope $\hmu$. 
The
shaded area is the transient information $\TI$.  For more discussion,
see text.} 
\label{HvsL}
\end{figure}
% **********************************************************************

The {\em source entropy rate} $\hmu$ is the rate of increase with
respect to $L$ of the total Shannon entropy in the large-$L$ limit:
\begin{equation}
    \hmu \equiv \lim_{L \rightarrow \infty} \frac{H(L)}{L} \; ,
\label{ent.def}
\end{equation}
where $\mu$ denotes the measure over infinite sequences that induces 
the $L$-block joint distribution ${\rm Pr} (s^L)$; the units are
{\em bits/symbol}.  Alternatively, one can define a finite-$L$
approximation to $h_\mu$, 
\begin{equation}
  h_\mu(L) \, = \, H(L) - H(L\!-\!1) \;,
\label{h.def}
\end{equation}
and show \cite{Cove91} that
$h_\mu = \lim_{L \rightarrow \infty} h_\mu(L)$. 
%Having implicitly discounted for the correlations
%and structures in longer and longer sequence blocks, 
The entropy rate $h_\mu$ quantifies the irreducible randomness in
observation sequences produced by the environment---the randomness
that persists even after statistics over longer and longer blocks of
observations are accounted for by the agent.  

% **********************************************************************

\section{Excess Entropy: Measuring Memory}

Having looked at length-$L$ sequences, an agent can estimate the 
true randomness $\hmu$ by calculating $h_\mu(L)$, defined in
Eq.~(\ref{h.def}).  With enough sensory data it can get good
approximations 
to $\hmu$ by using long sequences. But what if there is insufficient
data to allow this? To answer this we must determine how the estimates 
$h_\mu(L)$ converge  to $h_\mu$?  One measure of convergence is
provided by the {\em excess entropy} $\EE$:
\begin{equation}
\EE \, \equiv \, \sum_{L=1}^\infty [\hmu(L) - \hmu] \;,
\label{E.def}
\end{equation}
The units of $\EE$ are {\em bits}. The excess entropy is not a new
quantity; it was first introduced almost two decades ago
\cite{Crut83a,Gras86}. For recent reviews see
\cite{Crut01a,Shal98a,Bial00a}. 

$\EE$ measures the convergence of $h_\mu(L)$ and plays a role in
determining how an agent comes to know how random its environment
is. But what exactly does $\EE$ measure? The length-$L$ approximation 
$h_\mu (L)$ overestimates the entropy rate $h_\mu$ at finite $L$ by an
amount $h_\mu(L) - h_\mu$.  This difference measures how much more
random single measurements appear using the finite $L$-block
statistics than the statistics of infinite sequences. In other words,
this excess randomness tells us how much additional information must
be gained from the environment in order to reveal the actual per-symbol
uncertainty $h_\mu$. Thus, we can think of the difference $h_\mu (L) -
h_\mu$ as the redundancy (per symbol) in length-$L$ sequences: that
portion of information-carrying capacity in the $L$-blocks which is not
actually random, but is due instead to correlations. The excess entropy
$\EE$, then, is the total amount of this redundancy and, as such, a 
measure of one type of memory intrinsic to an environment.

The next proposition establishes a geometric interpretation of $\EE$
and an asymptotic form for $H(L)$.

\begin{proposition} The excess entropy is the subextensive part of
$H(L)$; that is,
\begin{equation}
  \EE = \lim_{L \rightarrow \infty} [ H(L) - \hmu L ]\;.
\end{equation}
\label{EEfromEntropyGrowth}
\end{proposition}

This proposition implies the following asymptotic form for $H(L)$: 
\begin{equation} 
  H(L) \, \sim \, \EE + \hmu L \,, \;{\rm as} \; L \rightarrow \infty \;.
\label{EEScalingForm}
\end{equation}
Thus, we see that $\EE$ is the $L = 0$ intercept of the linear
function Eq.~(\ref{EEScalingForm}) to which $H(L)$ asymptotes. This
observation, also made in
Refs.~\cite{Gras86,Bial00a,Shaw84,Li91}, is shown graphically
in Fig.~\ref{HvsL}.

Another way to understand excess entropy is through its expression
as a type of mutual information. 

\begin{proposition} The excess entropy is the mutual information
between the past and the future:    
\label{EandMI_Proposition}
\begin{equation}
  \EE \, = \,\lim_{L \rightarrow \infty}  I[s_0 s_1 \cdots s_{2L-1};
  s_{2L} s_{2L+1} s_{2L-1}] \;,
\label{EandMI}
\end{equation}
when the limit exists.
\end{proposition}

Eq.~(\ref{EandMI}) says that $\EE$ measures the amount of historical
information stored in the present that is communicated to the future.
For a discussion of some of the subtleties associated with this
interpretation, however, see Ref.~\cite{Shal98a}.
Prop.~\ref{EandMI_Proposition} also shows that $\EE$ can be
interpreted as the {\em cost of amnesia}: If an agent suddenly loses
track of its environment, so that it cannot be predicted at an error
level determined by the entropy rate $\hmu$, then the environment
appears more random by a total of $\EE$ bits.

% **********************************************************************
\section{Transient Information: A Measure of Synchronization}
\label{T_Section}
% **********************************************************************

We now introduce a quantity that answers the question, How does $H(L)$
converge to its asymptote $\EE + \hmu L$? That is, when has an agent
made a sufficient number of observations that it can determine the
complexity of its environment? The answer to these questions is provided
by the {\em transient information} $\TI$: 
\begin{equation}
\TI \equiv \sum_{L=0}^\infty \left[ \EE + \hmu L - H(L) \right] \;. 
\label{T.def}
\end{equation}
Note that the units of $\TI$ are {\em bits $\times$ symbols}.  The
transient information is a new quantity, recently introduced by us in
Ref.~\cite{Crut01a}. 

Thus, for finite-memory ($\EE$ and $\TI$ finite) processes $H(L)$ scales
as $\EE + h_\mu L$ for
large $L$. When this scaling form is attained, we say that the agent
is {\em synchronized} to the environment. In other words, when
\begin{equation}
\TI (L) \, \equiv \, \EE + \hmu L - H(L) \, = \, 0 \;,
\label{OSSyncCondition}
\end{equation}
we say the agent is synchronized at length-$L$ sequences. As we will
see below, agent-environment synchronization corresponds to the agent
being in a condition of knowledge such that it can predict the
environmental observations at an error rate commensurate with to the
environment's
entropy rate $\hmu$. We refer to $\TI$ as {\em transient} since
during synchronization the agent's prediction probabilities change,
stabilizing only after it has collected a sufficient number of
observations. 

To ground this interpretation, we can establish a direct relation
between the transient information $\TI$ and the amount of information
required for synchronization to block-Markovian environments.
Assume that the agent has a correct model ${\cal M}=\{{\cal V}, T\}$
of the environment, where ${\cal V}$ is a set of states and $T$ is the 
rule governing transitions between states.
The task for the agent is to make observations and determine the
state $v \in {\cal V}$ of the environment. Once the agent knows
with certainty the current state, it is {\em synchronized} to the
environment, and the average per-symbol uncertainty is exactly $h_\mu$.
We are interested in describing how difficult it is to synchronize to
a directly observed Markov process.   

The agent's knowledge of ${\cal V}$ is given by a distribution over
the states $v \in {\cal V}$. Let ${\rm Pr}( v | s^L, {\cal M} )$
denote the distribution over $\cal V$, given that the particular
sequence $s^L$ has been observed and the agent has internal model
${\cal M}$. The entropy of this distribution measures the agent's
average uncertainty in predicting $v \in {\cal V}$. Averaging this
uncertainty over the possible length-$L$ sequences, we obtain the
{\em average agent-environment uncertainty}: 
\begin{eqnarray}
  {\cal H}(L) \, \equiv & & \nonumber \\
  & & \hspace{-1.5cm} \, - \sum_{s^L} {\rm Pr}( s^L )\sum_{v \in
  {\cal V}} 
  {\rm Pr}( v | s^L, {\cal M}) \log_2  {\rm Pr}( v | s^L, {\cal M}) \;.
\label{script.H.def}
\end{eqnarray}
The quantity ${\cal H}(L)$ can be used as a criterion for
synchronization. The agent is synchronized to the environment when
${\cal H}(L) = 0$---that is, when the agent is completely
certain about the state $v \in \cal V$ of the mechanism generating
the sequence. When the condition in
Eq. (\ref{OSSyncCondition}) is met, we see that ${\cal H}(L) = 0$, and
the uncertainty associated with the prediction of the model ${\cal M}$
is exactly $h_\mu$. 

However, while the agent is still unsynchronized ${\cal H}(L) > 0$.
We refer to the total average uncertainty experienced by an
agent during the synchronization process as the {\em
synchronization information} $\SI$: 
\begin{equation}
\SI \equiv \sum_{L=0}^\infty {\cal H}(L) ~.
\end{equation}
The synchronization information measures the average total information
that must be extracted from observations so that the agent is
synchronized.   

In the following, we assume that the environment is Markovian of order
$R$. In contrast to the scenario depicted in
Fig.~\ref{MeasurementChannel}, we assume that the Markov model is not
hidden, in the sense that internal states are directly observable.

\begin{theorem}
\label{SyncTheorem}
For an order-$R$ Markovian environment, the synchronization information
$\SI$ is given by:
\begin{equation}
\SI = \TI + \frac{1}{2} R(R+1) \hmu \;.
\label{SyncTheoremEquation}
\end{equation}
\end{theorem}

Thus, the transient information $\TI$---together with the entropy rate
$h_\mu$ and the order $R$ of the Markov process---measures how difficult
it is to synchronize to an environment. If a system has a large $\TI$,
then, on average, an agent will be highly uncertain about the internal
state of the environment while synchronizing to it. Thus, $\TI$
measures a structural feature of the environment: how difficult it is
for an agent to synchronize to it.

%The transient information measures a structural property of the
%system---a property not captured by the excess entropy $\EE$.   

% **********************************************************************
\section{Applications and Implications}
\label{Applications}

Using $\hmu$, $\EE$, and $\TI$ one can distinguish various types of
entropy convergence and different structural classes of environment.
We can now return to the set of questions posed in the introduction:
How can we untangle different sources of apparent randomness? In
particular, what happens to estimates of the environment's randomness
if we ignore its structure? 

Here we show that there are direct and empirically important
consequences for ignoring structural properties. Namely, missed
regularities are converted to apparent randomness, assumed memory
produces false predictability, and assumed synchronization leads to
memory underestimates. These result in a range of misleading inferences
about both the environment's randomness and its structure. We consider
four different issues:
\begin{enumerate}
\item What happens when an agent ignores entropy-rate convergence?
\item What happens when the environment's apparent memory is ignored?
\item What happens if the agent ignores synchronization? 
\item What happens if the agent assumes it is synchronized to the
environment, when it is not?
\end{enumerate}

% **************************************************
\subsection{Disorder as the Price of Ignorance}

The first two questions are closely related and rather straightforward 
to answer.  The preceding sections defined several different
quantities---$\hmu$, $\EE$, and $\TI$---that measure
randomness, memory, synchronization, and other
features of a process. For the most part, these are asymptotic
quantities in the sense that they involve the behavior of the function
$H(L)$ in the $L \rightarrow \infty$ limit. Thus, their exact
empirical estimation demands that an infinite number of measurements
(for accurate estimates of sequence probabilities) of infinitely long
sequences be made.  Obviously, other than by analytic means, it is not
possible to calculate exactly such quantities. Exact, $L \rightarrow
\infty$ results are known for only a few special processes that are
analytically tractable.

This leads one to ask, Even when sequence probabilities are accurately
known, how well can these various environment properties be estimated
at finite $L$?  What errors are introduced, and are these errors
related in any way? 

The simplest such question, the first one listed above, arises when
one attempts to estimate source randomness $h_\mu$ via the
approximation $h_\mu(L)$. Generally, stopping the estimate at finite
$L$ gives one a rate $\hmu(L)$ which is larger than the actual rate
$\hmu$.  That is, the environment appears {\em more random} if we
ignore correlations between observations separated by more than $L$
steps.  For a discussion of several methods to improve on the
estimator $h_\mu(L)$ in the context of dynamical systems, see
Ref.~\cite{Schu96}.

%This follows directly from the definitions of $h_\mu$ and
%$h_\mu(L)$.  However, it turns out that this form of overestimation
%of $h_\mu$ is related to the excess entropy $\EE$.
%There is a quantitative trade-off between randomness and memory.

% **********************************************************************
\begin{figure}[tbp]
%\vspace{-0.2in}
%\hspace{0.2in}
\epsfxsize=3.0in
\begin{center}
\leavevmode
\epsffile{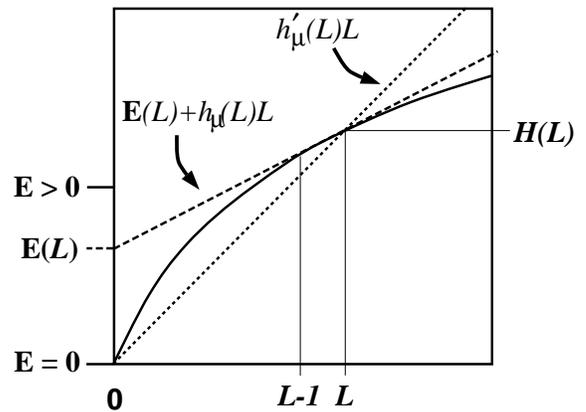}
\end{center}
\vspace{2mm}
\caption{Ignored memory is converted to randomness: Illustration of how
ignoring memory, in this case implicitly assuming $\EE = 0$ as
Eq.~(\ref{ent.def}) implies, when actually $\EE > 0$, leads to an
overestimate $\hmu^\prime (L)$ (slope of dotted line) of the actual
entropy rate $\hmu$ (slope of dashed line).} 
\label{MemoryToDisoder}
\end{figure}
% **********************************************************************

An agent could also estimate $h_\mu$ at finite $L$ by using 
$\hmu^\prime (L) = H(L)/L$, as suggested by Eq.~(\ref{ent.def}).
Using this definition to estimate $\hmu$ is tantamount to assuming
that $\EE = 0$, as illustrated by the dashed line $\hmu^\prime L$ in
Fig.~\ref{HvsL}. Now suppose an agent makes measurements of an
environment with entropy rate $h_\mu$ and excess entropy $\EE >0$.
Then, at a given $L$, we can ask what the entropy rate estimate
$\hmu^\prime (L) = H(L)/L$ is.  As shown in \cite{Crut01a},
$\hmu^\prime (L) \geq h_\mu(L) \geq \hmu$.  But how much more random
does the environment appear? This is answered in a straightforward way
by the following proposition.

\begin{proposition}
When the agent is synchronized to the environment,
\begin{equation}
h_\mu^\prime (L) - \hmu = \frac{\EE}{L} ~.
\end{equation}
\label{StructureToRandom}
\end{proposition}

In this way, $\EE$ bits of memory are converted into additional,
apparent randomness. The environment appears more random due to the
agent's ignoring one of its structural properties.

Although $\EE$ is an $L$-asymptotic quantity, the error $\EE/L$ in the
entropy-rate estimate dominates at small $L$. Moreover, being
restricted to small $L$ is typical of experimental situations with
limited data or in which drift is present. One cannot
reliably estimate the $L$-block probabilities $\Pr (s^L)$ at large $L$
due to the exponential growth in their number or the nonstationarity
of block probabilities, respectively.

% **************************************************
\subsection{Predictability and Instantaneous Synchronization}

Conversely, if one assumes a fixed amount of memory $\EE$, we shall
see that this leads to an underestimate of the entropy rate $h_\mu$
and the environment appears more predictable than it is. 
Assuming a fixed excess entropy is not something that one is
likely to do in the particular setting here, in which an agent
empirically measures entropy density and related quantities from
observation sequences.  In a more general modeling setting, however,
one always runs the risk of using too large a model and, in so doing,
``projecting'' some particular structure---such as, additional
memory capacity---onto the environment. Assuming a
fixed, nonzero value for the excess entropy is, in an abstract sense,
an example of over-fitting.  Given this, we ask, What is the
consequence of assuming a fixed value for $\EE$? 

\begin{figure}[tbp]
%\vspace{-0.2in}
%\hspace{0.2in}
\epsfxsize=3.0in
\begin{center}
\leavevmode
\epsffile{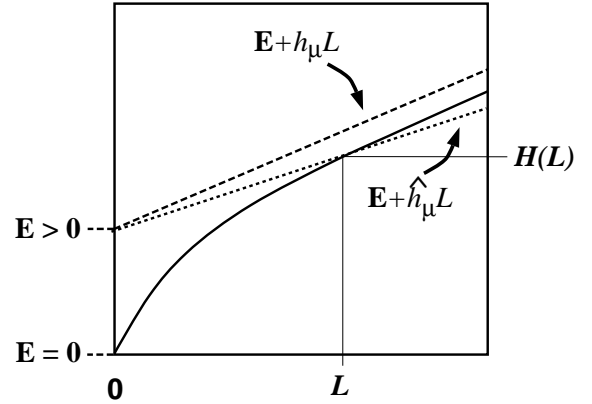}
\end{center}
\vspace{2mm}
\caption{Assumed synchronization converted to false predictability:
  Schematic illustration of how assuming one is synchronized, leads
  to an underestimate $\widehat{\hmu}$ (slope of dotted line) for an
  environment with excess entropy $\EE > 0$ and entropy rate $\hmu$
  (slope of dashed line).} 
\label{InstantSyncOrder}
\end{figure}

Equivalently, what happens if the agent assumes that it is synchronized
to the environment at some finite $L$, implying that
$H(L) = \EE + h_\mu L$ at that $L$?  The geometric construction for
this scenario
is given in Fig.~\ref{InstantSyncOrder}.  In effect the environment is
erroneously considered to be a completely observable Markovian
process in which $H(L)$ converges to its asymptotic
form exactly at some finite $L$ \cite{Crut01a,Ebel97b}.  
If the agent then uses its assumed
value for $\EE$, one arrives at the estimator $\widehat{\hmu}$ where
\begin{equation}
\widehat{\hmu} \, \equiv \, \frac{H(L) - \EE}{L} \, \neq \hmu \;.
\label{widehat.hmu.def}
\end{equation} 
At a given $L$ the effect is
that the agent considers the environment to have a larger $\EE$ than it
actually has at that $L$. The line $\EE + \widehat{\hmu} L$ appears
fixed at $\EE$ when that intercept should be lower at the
given $L$. The result, easily
gleaned from Fig.~\ref{InstantSyncOrder}, is that the entropy rate
$\hmu$ is underestimated as $\widehat{\hmu}$.  (The two entropy rates
are the slopes of the two straight lines.)  In other words, the agent
will believe the environment to be more predictable than it actually
is.  

\begin{proposition}$\;$
An agent monitors an environment with excess entropy $\EE > 0$. If the
agent assumes it is synchronized when it is not, then
\begin{equation}
\widehat{\hmu} \leq \hmu \;.
\end{equation}
\end{proposition}

% **********************************************************************
% **********************************************************************

% **************************************************
\subsection{Assumed Synchronization Implies Reduced Apparent Memory}

\begin{figure}[tbp]
%\vspace{-0.2in}
%\hspace{0.2in}
\epsfxsize=3.0in
\begin{center}
\leavevmode
\epsffile{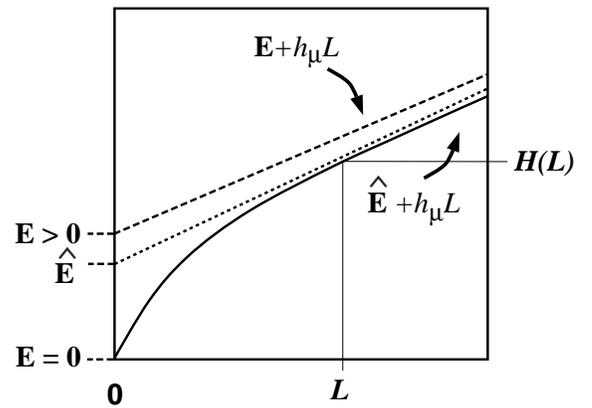}
\end{center}
\vspace{2mm}
\caption{Assumed synchronization leads to less apparent memory:
Schematic illustration of how assuming synchronization to an
environment, in
this case implicitly assuming $H(L) = \EE + \hmu L$, leads to an
underestimate $\widehat{\EE}$ of the actual memory $\EE > 0$.}
\label{SyncLessMemory}
\end{figure}

In addition to analyzing the effects on the apparent entropy rate due
to assuming synchronization, we can ask a complementary question:
What are the effects of assuming synchronization on estimates
$\widehat{\EE}$ of the apparent memory? Figure \ref{SyncLessMemory}
illustrates this situation. 

% **********************************************************************
% **********************************************************************

If, at a given $L$, we approximate the entropy rate estimate
$\hmu(L) = H(L) - H(L-1)$ by the true entropy $\hmu$, then the offset
between the asymptote and $H(L)$ is simply $\EE + \hmu L - H(L)$.
Thus, looking at Fig.~\ref{SyncLessMemory}, we see that we have a
reduced apparent memory $\widehat{\EE} \leq \EE$ of 
\begin{equation}
\widehat{\EE} = H(L) - \hmu L ~.
\end{equation}
In fact, since the estimated entropy rate is larger than $\hmu$, the
reduction in apparent memory is even larger. Thus, assuming
synchronization, in the sense that $h_\mu(L) = h_\mu$, leads one to
underestimate the apparent memory, as measured by the excess entropy
$\EE$. And so, the environment appears less structurally complex
than it is.

% **********************************************************************
\section{Conclusion}
\label{Conclusions}

We have reviewed several information theoretic measures of an
environment's randomness and several of its structural properties. We
also introduced a new quantity, the transient information $\TI$. One
of the central results of this work is contained in Theorem
\ref{SyncTheorem}, which states that $\TI$ is directly related to the
total agent-environment uncertainty experienced while an agent
synchronizes to a Markovian environment.    

We then considered various trade-offs between finite-$L$ estimates of
the excess entropy $\EE$ and the entropy rate $h_\mu$. In particular,
we showed that if an agent does not take one or another into account it
will systematically {\em over-} or {\em underestimate} an environment's
entropy rate $\hmu$.  For example, there can be an inadvertent
conversion of ignored memory into apparent randomness. The magnitude oa
f this effect is proportional to the difference between environment
memory and the upper bound on memory that the agent store in its
internal model. In a complementary way, one can inadvertently
convert assumed memory into false predictability.  As a result, an
agent must have some method for accounting for an environment's
structural features, even if it's focus is only on the apparently
simple question of how random a process is \cite{Ford83}.

\section*{Acknowledgments}

This work was supported at the Santa Fe Institute under the
Computation, Dynamics, and Inference Program via SFI's core grants
from the National Science and MacArthur Foundations. Direct support
was provided from DARPA contract F30602-00-2-0583.

% **************************** REFERENCES ****************************

\bibliography{spin}

\end{multicols}

\end{document}